\documentclass[fleqn]{llncs}
\usepackage{version}
\usepackage{os2}

\title{On the Definition of a Theoretical Concept of an Operating System}
\author{J.A. Bergstra \and C.A. Middelburg}
\institute{Informatics Institute, Faculty of Science,
           University of Amsterdam, \\
           Science Park~107, 1098~XG Amsterdam, the Netherlands \\
           \email{J.A.Bergstra@uva.nl,C.A.Middelburg@uva.nl}}

\begin{document}
\maketitle

\begin{abstract}
% 44 %
We dwell on how a definition of a theoretical concept of an operating
system, suitable to be incorporated in a mathematical theory
\linebreak[2] of operating systems, could look like.
This is considered a valuable preparation for the development of a
mathematical theory of operating systems.
\end{abstract}

% operating system, theoretical concept, definition, theory, concurrency
% D.4.0

\section{Introduction}
\label{sect-introduction}

Presently, operating systems are a hot topic in the sector of
information and communication technologies.
General-purpose operating systems that have been developed for desktop
computers or laptop computers are not suitable for mobile devices, such
as smartphones, personal digital assistants, personal navigation devices
and e-book readers, due to the special needs of these devices, such as
regulation of power consumption to prolong battery life and real-time
responses for time-critical applications.
Therefore, the increasing importance of mobile devices has triggered the
development of many mobile operating systems.
There is really a very strong competition going on among various major
companies from the sector of information and communication technologies
in a bid for the most successful mobile operating system
(cf.~\cite{HO09a}).

We expect that a theoretical understanding of the concept of an
operating system will become increasingly important to the development
of successful operating systems.
However, it happens that in computer science since the introduction of
the first operating systems more than fifty years ago no serious
attention has been paid to the clarification of what is an operating
system.
Almost any explanation of the concept fails to capture the concept of an
operating system satisfactorily.
The existing theoretical understanding related to operating systems
concerns details of the functioning of operating systems, such as
scheduling the programs in execution and allocating resources to the
programs in execution, and shows little coherence.

We became fully aware of the state of affairs outlined above only after
an extensive search for publications on operating systems recently
carried out by one of us, which is reported on in~\cite{Mid10a}.
This state of affairs forms our motivation to have a closer look at the
concept of an operating system.
In this note, we dwell on how a definition of a theoretical concept of
an operating system, suitable to be incorporated in a mathematical
theory of operating systems, could look like.
This is considered a valuable preparation for the development of a
mathematical theory of operating systems.
We also make an effort to explain the circumstances which justify the
interest in defining a theoretical concept of an operating system.

First, we sketch how the concept of an operating system is dealt with in
publications on operating systems and give an explanation of the concept
distilled from statements about operating systems found in publications.
Next, we make some remarks about theoretical concepts that have come
into being as formalized versions of pragmatic concepts and present some
highlights of an elementary meta-theory about definitions of theoretical
concepts.
After that, we use the foregoing to outline how a definition of a
theoretical concept of an operating system could look like.

\section{The Concept of an Operating System in the Literature}
\label{sect-summary-outcome}

Recently, we have searched for publications in which one can find
reasons for introducing operating systems, statements that explain the
concept of an operating system, a definition of a formalized version of
the pragmatic concept of an operating system or a theory of operating
systems based on such a definition.
It turned out that the number of such publications is very small.
On the outcome of the search in question is extensively reported
in~\cite{Mid10a}.
Below, we give a brief summary of the outcome of this search.

It is often stated that Strachey's article on multiprogramming operating
systems from 1959~\cite{Str59a} is the first important article on
operating systems.%
\footnote
{Strachey's article can only be obtained by ordering a hard copy at the
 National Archive of the United Kingdom.}
It is quite surprising that the article of Codd and others on
multiprogramming operating systems from 1959~\cite{CLMS59a} is never
mentioned as the first important article on operating systems.
In the latter paper, motivation for, requirements for, and functions of
a multiprogramming operating system are given.
This can be taken for a preparation to the formulation of the scheduling
problem in multiprogramming operating systems in~\cite{Cod60a} and the
description of a scheduling algorithm for a multiprogramming operating
system in~\cite{Cod60b}.
Probably the article of Codd and others is as significant as the article
of Strachey.

Apart from the attempt of Codd and others in the above-mentioned
article, few serious attempts have been made to explain the concept of
an operating system; and apart from the reasons given by Codd and others
in the same paper, reasons for introducing operating systems are seldom
given.
Dennis and Van Horn make a serious attempt to explain the concept of an
operating system in~\cite{DH66a} and Denning makes another serious
attempt in~\cite{Den71a}, but most other attempts cannot be called
serious.
Examples of non-serious attempts are one-liners like ``an operating
system is an extended machine and a resource manager'' and enumerations
of the usual terms for the basic constituents of an operating system.
Cloot gives good reasons for introducing operating systems
in~\cite{Clo65a}, an article whose sole aim is to explain why the need
for operating systems has arisen, but usually the reasons are not more
advanced than ``it is useful to have an operating system available''.

In~\cite{YLSL99a}, Yates and others give an abstract model of an
operating system, using input/output automata, which could be used as a
starting point for the definition of a formalized version of the
pragmatic concept of an operating system.%
\footnote
{The article of Yates and others actually gives two models.
 The abstract model is the model that is called the user level model in
 the article.
}
However, that model is still not strong in capturing the pragmatic
concept of an operating system.
Apart from this, publications in which abstract models of an operating
system are given that could be used as a starting point for the
definition of a formalized version of the pragmatic concept of an
operating system are virtually absent.
Publications on theory about operating systems themselves are totally
absent.
In publications on operating systems that are of a theoretical nature,
one finds only theory about details of the functioning of operating
systems, such as scheduling the programs in execution and allocating
resources to the programs in execution.

From the outcome of the search, we conclude that the operating systems
community pays little attention to clarifying adequately what is an
operating system.
It happens that most publications on operating systems mainly concern
the following:
\begin{itemize}
\item
principles of operating system design;
\item
theory and techniques related to details of the functioning of operating
systems such as scheduling and resource allocation;
\item
issues concerning operating systems for multi-processor computers and
operating systems for networks of distributed computers;
\item
operating system support for security, privacy, fault-tolerance,
multi-media applications, et cetera;
\item
designs of, analyses of, and experiences with specific operating
systems.
\end{itemize}
It is striking that most of these publications give little insight in
the concept of an operating system.
Virtually all exceptions are articles published before 1970.
Our findings of the search agree with the findings of the study of
courses and textbooks presented in~\cite{CS00a}.

\section{An Explanation of the Concept of an Operating System}
\label{sect-added-bonuses}

During the search for publications on operating system, many statements
about operating systems were found from which we could distill the
explanation of the concept of an operating system given below.

An operating system is a system that provides a convenient execution
environment for programs that allows for multiple programs with shared
resources to be executed concurrently.
An operating system is responsible for:
\begin{enumerate}
\item
loading programs and starting their execution;
\item
scheduling the programs in execution;
\item
allocating resources to the programs in execution;
\item
preventing interference between the programs in execution;
\item
controlling the use of main memory by the programs in execution;
\item
storing and retrieving data organized into files and directories on
secondary storage devices;
\item
receiving data from input devices and sending data to output devices;
\item
communicating data over computer networks;
\item
controlling peripheral devices.
\end{enumerate}

It is customary to distinguish the following basic constituents in an
operating system:
\begin{itemize}
\item
process management, responsible for 1, 2, 3 and 4;
\item
memory management, responsible for 5;
\item
file management, responsible for 6;
\item
input/output management, responsible for 7;
\item
network management, responsible for 8;
\item
device management, responsible for 9.
\end{itemize}
Process management and a part of memory management are needed to provide
an execution environment for programs that allows for multiple programs
with shared resources to be executed concurrently.
Device management, network management, input/output management, file
management, and a part of memory management are needed to provide a
\emph{convenient} execution environment, because they hide interrupts,
networking protocols, device-dependent input, output and storage,
physical memory size, et cetera.

Operating systems can be classified as:
\begin{itemize}
\item
single-user or multi-user;
\item
non-interactive or interactive;
\item
single-tasking, non-preemptive multi-tasking or preemptive
multi-tasking.
\end{itemize}
Actually, the explanation given above is an explanation of the concept
of an multi-tasking operating system.
Single-tasking operating systems are border cases of operating systems:
the maximal number of programs that can be executed concurrently is only
one.
Clearly, a multi-tasking operating system is a more general concept than
a single-tasking operating system.
Batch operating systems, of which the first became probably operational
in 1956 (see~\cite{Ryc83a}), are multi-user, non-interactive,
single-tasking operating systems.
Multiprogramming operating systems, of which the first was probably
developed over the period 1957--1961 (see~\cite{KPH61a}), are
multi-user, non-interactive, (non-preemptive or preemptive)
multi-tasking operating systems.
Time-sharing operating systems, of which the first was probably
developed over the period 1961--1963 (see~\cite{CMD62a}), are
multi-user, interactive, preemptive multi-tasking operating systems.

The explanation given above has been obtained by extracting the essence
of many statements found in publications on operating systems.
By no means, we consider it an explanation that captures the concept of
an operating system satisfactorily.
However, at least it provides a reasonable picture of how is generally
thought about operating systems in the operating systems community.

\section{On the Definition of Computer Science Concepts}
\label{sect-definitions}

In order to make the answer on the question ``what is an operating
system?'' precise, we need an elementary meta-theory about answers on
questions of the form ``what is \ldots?'', i.e.\ an elementary
meta-theory about definitions.
Because we are interested in definitions that can be incorporated in
mathematical theories, the scope can be restricted to definitions of
theoretical concepts.
Below we present some highlights of an elementary meta-theory about
definitions of theoretical concepts.
Preceding that, we make some remarks about theoretical concepts that
have come into being as formalized versions of pragmatic concepts.

Any formalized version of a pragmatic computer science concept, such as
the concept of an operating system, differs from the informal one: it is
theoretical instead of pragmatic.
The difference is unavoidable because the formalized version is a
mathematical representation of the informal version.
It means that the instances of a pragmatic computer science concept
recognized as such in practice are not the same as the instances of its
formalized version considered in a theory based on the formalized
version.
Moreover, it is natural that the definition of a formalized version of a
pragmatic computer science concept brings about that not all instances
of the pragmatic concept are covered.
All this is certainly not specific to pragmatic computer science
concepts.
Similar remarks can be made with respect to many other concepts.
For example, the formalized version of the concept of a tree from graph
theory is definitely quite different from the informal one from botany.

What we consider an important property of a definition of a theoretical
concept is its bareness.
This means that it should be deprived of connotations concerning
secondary matters such as the purpose of instances of the concept, the
circumstances in which instances of the concept play a role, and the
dependencies between instances of the concept and instances of another
concept that are not conceptual.
For example, a bare definition of a theoretical concept of a program
does not have connotations such as ``the purpose of a program is to
produce a certain behaviour'', ``a program plays a role in the case
where a behaviour is produced by means of a computer'', and ``a program
depends on a computer in order to be executed''.

A conceptual dependency is made apparent in a definition of a
theoretical concept if the concept in question is defined in terms of
another theoretical concept.
Conceptual dependencies made apparent in a definition do not decrease
its bareness.
In a family of concepts which are somehow connected by conceptual
dependencies, some concepts may be more central than others.
For example, a theoretical concept of a program, a theoretical concept
of a machine and a theoretical concept of a run of a program on a
machine might form a family of concepts where the concept of a run is
conceptually dependent of the other two concepts, and the concept of a
program is most central.

In the case of such a family of concepts, it seems useful to consider
the collection of definitions of all concepts in the family, together
with a stratification indicating how central each of the concepts is, as
a whole.
We coin the term stratified concept family definition for such a whole.
Of course, the concept definitions in a stratified concept family
definition should be bare definitions.
Although many mathematical theories are build on a stratified concept
family definition, we could not find any meta-theory of definitions
covering something like stratified concept family definitions with the
exception of the meta-theory of definitions presented in~\cite{FF90a}.
Stratified concept family definitions resemble the definition dags
introduced in that paper.

To accommodate various kinds of utility and value analysis, it appears
to be useful to extend a stratified concept family definition with
definitions of measures that represent the utility or value of instances
of the different concepts involved or groups thereof.
An alternative is to regard such measures as additional concepts which
are less central than all other concepts in the family in question.

\section{The Definition of the Concept of an Operating System}
\label{sect-definition-os}

Below, we outline how a definition of a theoretical concept of an
operating system could look like.
For that, we make use of the highlights of an elementary meta-theory
about definitions of theoretical concepts presented above.

A theoretical concept of an operating system is a formalized version
of the pragmatic concept of an operating system.
This implies that its definition is an explicative definition, which is
adequate for certain purposes and/or in certain contexts only.
To be able to connect a theory about operating systems to a large part
of the literature on operating systems, we therefore do not exclude the
possibility that the theory will include definitions of different
theoretical concepts of an operating system.
In what follows, we will not pay attention to this possibility and use
the phrase ``\emph{the} theoretical concept of an operating system''.

We know from the search for publications on operating systems mentioned
before that the ambition to give a definition of the theoretical concept
of an operating system is new.
We believe that a bare definition of the theoretical concept of an
operating system is possible.
Our starting-point for such a definition is the perception of an
operating system as a component of an analytic execution architecture
for programs as described in~\cite{BP04a} enriched by mechanisms by
which a program can switch over execution to another program and
interrupt the execution of another program.
Therefore, we think that the theoretical concept of an operating system
is at least conceptually dependent of a theoretical concept of a program
and a theoretical concept of an analytic execution architecture.
\linebreak[2]
From the definitions of these three concepts, we can put together a
stratified concept family definition where the theoretical concept of an
operating system is most central.
Such a stratified concept family definition provides a rationale for the
technicalities of the definition of the theoretical concept of an
operating system.

The above-mentioned mechanisms for program execution switch-over and
interruption give rise to a form of interleaving.
This means that the theory to be developed needs a concurrency theory as
a basis.
The question is what is a suitable underlying concurrency theory.
We expect that a suitable underlying concurrency theory can be obtained
by extending the concurrency theory developed in~\cite{BM08f}, which
covers program execution switch-over but does not cover program
execution interruption.
In the case where the underlying concurrency theory is obtained thus,
the analytic execution architectures involved in the definition of the
theoretical concept of an operating system are quite similar to the ones
discussed in~\cite{BM08f}.
They include a collection of programs between which execution can be
switched.
One of the programs in the collection is the operating system and the
others are the programs whose concurrent execution is controlled by the
operating system.
No matter what underlying concurrency theory is taken, it will introduce
additional theoretical concepts of which the theoretical concept of an
operating system is conceptually dependent.

\section{Concluding remarks}
\label{sect-conclusions}

As a preparation for the development of a mathematical theory of
operating systems, we have dwelled on how a definition of a theoretical
concept of an operating system could look like.
In doing so, we were led to present some highlights of an elementary
meta-theory about definitions of theoretical concepts.
We believe that such a meta-theory has wider applicability and deserves
further elaboration.

\bibliographystyle{splncs03}
\bibliography{OS}

% \par \vfill \par \noindent DRAFT of \today

\begin{comment}
... and how could be dealt with the potential problem that there exist
multiple conflicting concepts of an operating system.

Initially, the lack of a suitable theory of concurrency has been
responsible for the lack of a precise definition of the concept of an
operating system.

We think that the abstraction level offered by an operating system may
be lower than the abstraction level offered by the underlying machine.
\end{comment}

\end{document}